\documentclass{osa-article}

\journal{oe}

\begin{document}

\title{High Kerr nonlinearity of water in THz spectral range}

\author{A.N. Tcypkin,\authormark{1,*} M.V. Melnik,\authormark{1} M.O. Zhukova,\authormark{1} I.O. Vorontsova,\authormark{1} S.E. Putilin,\authormark{1} S.A. Kozlov,\authormark{1} and X.-C. Zhang\authormark{1,2}}

\address{\authormark{1}International Laboratory of Femtosecond Optics and Femtotechnologies, ITMO University, St. Petersburg, 197101, Russia\\
\authormark{2}The Institute of Optics, University of Rochester, Rochester, NY 14627, USA}

\email{\authormark{*}tsypkinan@corp.ifmo.ru} 



\begin{abstract}
The values of the nonlinear refractive index coefficient for various materials in the terahertz frequency range exceed the ones in both visible and NIR ranges by several orders of magnitude. We report the direct measurement of the nonlinear refractive index coefficient of liquid water by using the Z-scan method with broadband pulsed THz beam. Our experimental result shows that nonlinear refractive index coefficient in water is positive and can be as large as 3.5$\times$10$^{-10}$ cm$^{2}$/W in the THz frequency range, which exceeds the ones for the visible and NIR ranges by 6 orders of magnitude.
\end{abstract}

\section{Introduction}
Terahertz (THz) frequency range is finding more and more applications in different fields of fundamental research \cite{baierl2016nonlinear} and a variety of everyday applications such as nondestructive spectroscopy \cite{dong2017global} and imaging \cite{kaltenecker2016ultrabroadband}, communications \cite{nagatsuma2016advances, grachev2018wireless} and ultrafast control \cite{bowlan2017probing} as well as biomedicine \cite{smolyanskaya2018glycerol}. Recent advances in research have brought high intensity broadband sources of THz radiation into play \cite{zhang2017extreme}. These perspectives have already been shown for highly intensive THz pulses generated from organic crystals with peak intensity $I_{peak}$ $\sim$ 1.1$\times$10$^{13}$ W/cm$^{2}$  \cite{shalaby2015demonstration}. The exploration of matter nonlinearities in THz frequency region can open new directions for devices and systems development, components and instrumentation operation.

Despite the growing interest to this issue, the observations of nonlinearities in THz region were carried out without using direct measurements of material properties in most cases. The latter ones can be nonlinearity and dispersion in the wave propagation, nonlinear optical response caused by intense ultrashort THz pulses \cite{gaal2006nonlinear}, absorption bleaching by means of pump-probe technique \cite{hebling2010observation, hoffmann2009thz}, nonlinear free-carrier response \cite{sharma2012carrier, turchinovich2012self}, field-induced transparency \cite{li2015terahertz}, spin response \cite{baierl2016terahertz}, giant cross phase modulation and THz-induced spectral broadening of femtosecond pulses \cite{vicario2017subcycle}, quadratic THz optical nonlinearities by measuring the quadratic THz Kerr effect \cite{lin2018measurement}.

The most important parameter characterizing the nonlinearity of the material response in the field of intense waves is the coefficient of its nonlinear refractive index, usually denoted as $n_2$. In the first report of such kind of measurements \cite{kaur2010terahertz} silicon was tested by use of the open aperture Z-scan technique. Originally, it was predicted theoretically \cite{dolgaleva2015prediction} that nonlinear refractive index coefficient for various materials in the THz frequency range exceeds the ones in both visible and NIR ranges by several orders of magnitude. Then this fact was proven experimentally by the Z-scan method \cite{woldegeorgis2018thz, Tcypkin2018}.  However, in these works only indirect estimations of $n_2$ were made. 

Until recently, the study of water in the THz range has been considered impossible due to its large absorption. The broadband THz wave generation from water film \cite{jin2017observation, doi:10.1063/1.5054599} was experimentally demonstrated and opened a new field of interest. In this article, we present the direct measurement of water nonlinear refractive index coefficient for broadband pulsed THz radiation with the conventional Z-scan method. Since the Z-scan method works only with plane-parallel samples, we use flat water jet. We demonstrate that the nonlinear refractive index exceeds its visible and NIR ranges' values by 6 orders of magnitude \cite{paillette1969recherches, smith1977superbroadening, ho1979optical,boyd2003nonlinear}.

\section{Experiment and analytical model}
\begin{figure}[htbp]
\centering
\includegraphics[width=10 cm]{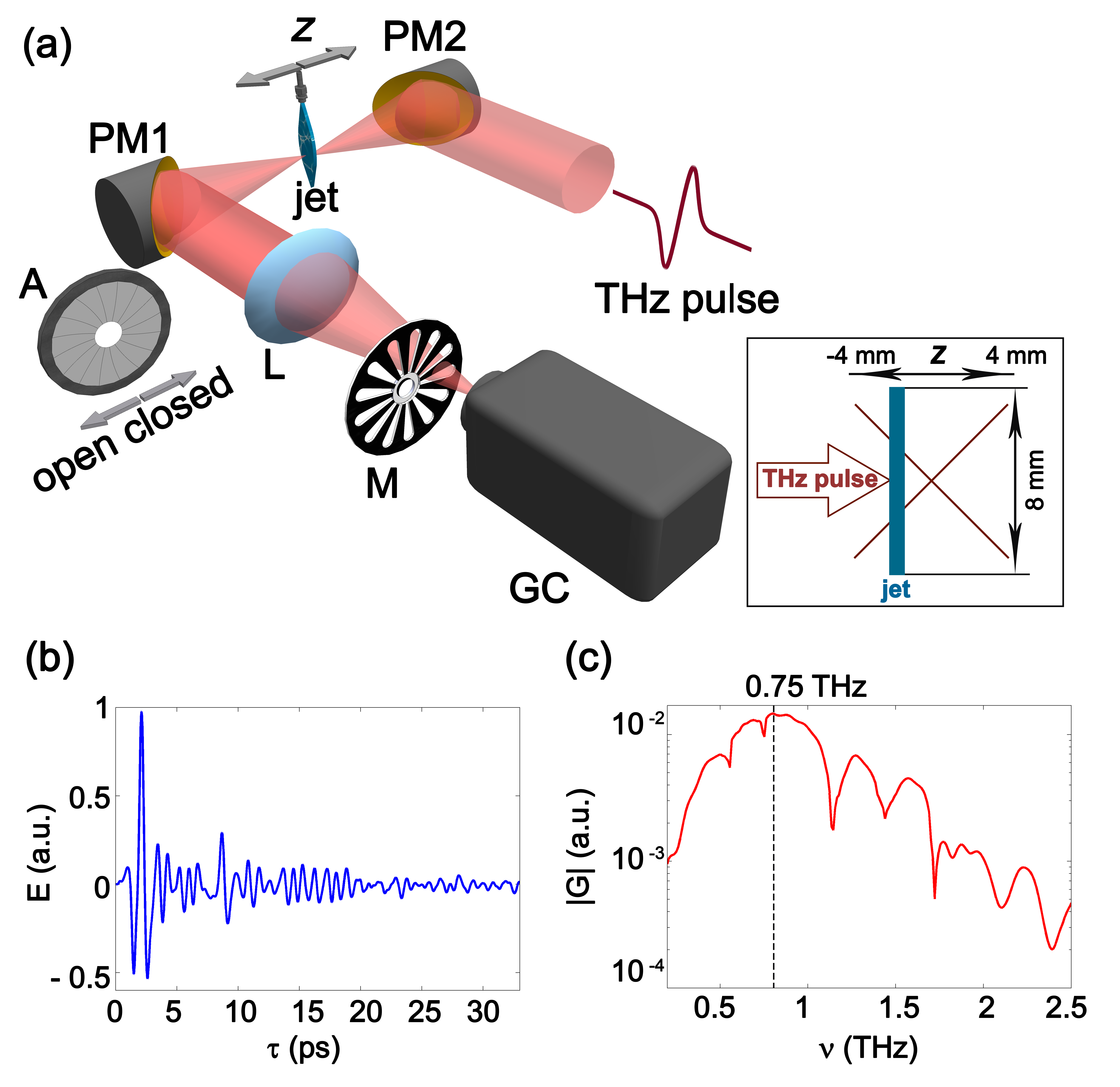}
\caption{(a) The experimental setup for measuring the nonlinear refractive index ($n_2$) of a liquid jet in the THz spectral range. Two parabolic mirrors (PM1 and PM2) with a focal length of 12.5 mm form the caustics area where the water jet (jet) is scanned along the z axis. The synchronization is performed using the mechanical modulator (M) located between the lens and the Golay cell (GC). The aperture (A) is moved from open to closed position to change the geometry of Z-scan from open to closed aperture. Insert - Geometrical position of the jet moved along the z axis relative to the THz radiation. The temporal waveform  (b) and its spectrum (c) of the THz pulse generated by the TERA-AX system.}
\label{fig1}
\end{figure}

Fig.\ref{fig1}(a) illustrates an experimental setup for measuring nonlinear refractive index (n$_2$) of a flat liquid jet with THz pulses. We use TERA-AX (Avesta Project) as a source of THz radiation. In this system the generation of THz radiation is based on the optical rectification of femtosecond pulses in a lithium niobate crystal \cite{yang1971generation}. The generator TERA-AX is pumped using a femtosecond laser system (duration 30 fs, pulse energy 2.2 mJ, repetition rate 1 kHz, central wavelength 800 nm). The THz pulse energy is 400 nJ, the pulse duration is 0.5 ps (Fig.\ref{fig1}(b)) and the spectrum width from 0.1 to 2.5 THz (Fig.\ref{fig1}(c)). The THz radiation intensity is controlled by reducing the femtosecond pump beam intensity. Pulsed THz radiation is focused and collimated by two parabolic mirrors (PM1 and PM2) with a focal length of 12.5 mm. The spatial size of the THz radiation at the output of the generator is 25.4 mm. Caustic diameter is 1 mm (FWHM).  In order to get a higher intensity in the waist, we use a short-focus parabolic mirror with a large NA. This geometry allows us to get the radiation intensity in the caustic of the THz beam 10$^8$ W/cm$^{2}$. Flat water jet (jet) is moved along the caustic area from -4 mm to 4 mm using a motorized linear translator, the restriction on the displacement is determined by the jet width and the focusing geometry of the THz radiation (see Fig.\ref{fig1}(a) insert). The polarization of the THz radiation is vertical. In this experiment we used distilled water which does not contain any substances as medium. The water jet has a thickness of 0.1 mm and is oriented along the normal to the incident radiation. The jet is obtained using the nozzle which combines the compressed-tube nozzle and two razor blades \cite{watanabe1989new}. This design forms a flat water surface with a laminar flow. The optical path of the THz pulse passes through the center of the jet area with a constant thickness. Thanks to the use of the pump water is released under the pressure. The hydroaccumulator in the system of water supply allows to significantly reduce the pulsations associated with the operation of the water pump. The THz radiation is collimated by the parabolic mirror PM2 and focused by the lens (L) on Golay cell (GC). For the closed aperture geometry the aperture (A) is moved in the beam (closed position). The synchronization is performed using the mechanical modulator (M) located between the lens and the Golay cell. When the jet is moved along the z axis through the focal region of the THz radiation, the average power of the THz radiation is measured using open and closed aperture.

\begin{figure}[htbp]
\centering
\includegraphics[width=14 cm]{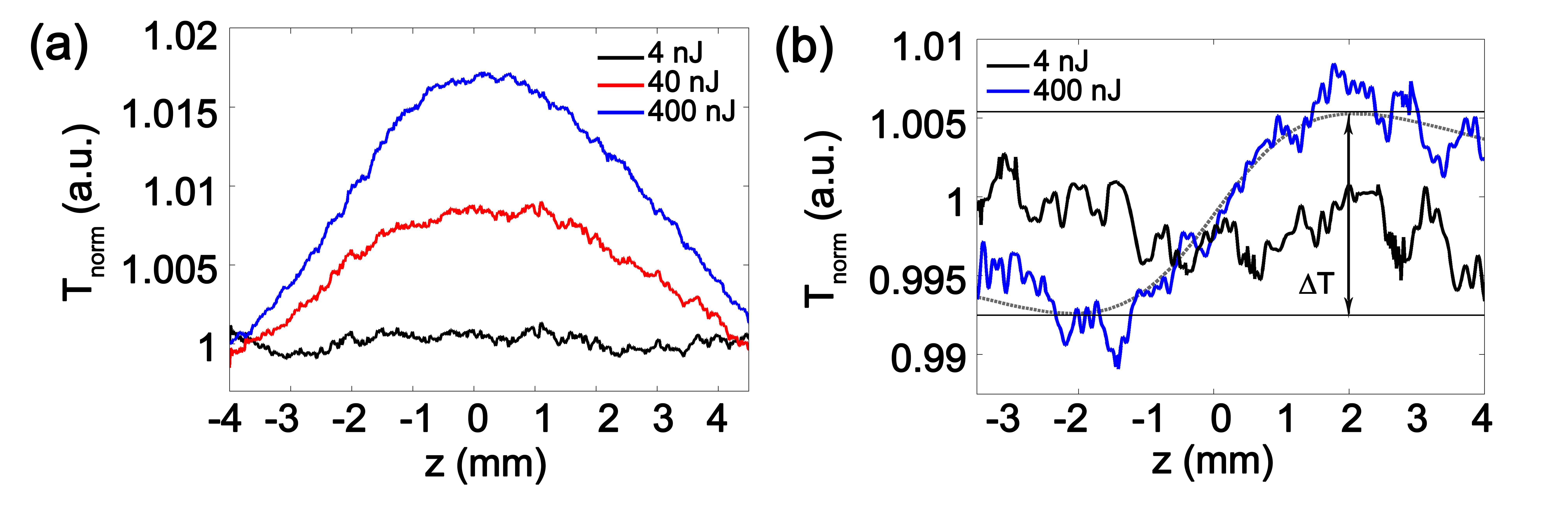}
\caption{Z-scan curves for a 0.1 mm thick water jet measured with open (a) and closed (b) aperture for different THz radiation energy values of 4 nJ, 40 nJ and 400 nJ. $\Delta$ T = 0.013 is the differential of the Z-scan curve measured with the closed aperture of radius 1.5 mm.}
\label{fig2}
\end{figure}

Despite the fact the experimental setup implies nonparaxial radiation, as was shown in \cite{kislin2018self} for pulses from a small number of oscillations, the differences between the paraxial and nonparaxial modes are negligible. Fig.\ref{fig2} shows Z-scan curves for the water jet measured with open (a) and closed (b) aperture for different values of the THz radiation energy. Each line is averaged over 50 measurements. (Fig.\ref{fig2}(a)) shows the water bleaching by around 2\% which is caused by THz radiation pump energy growth by 2 orders. For n$_2$ determination we use experimental data with the closed aperture (Fig.\ref{fig2}(b)).

Usually, Z-scan technique is strictly valid only for quasi-monochromatic radiation. However, it is also widely used in the case of femtosecond pulsed radiation which spectrum is quite large \cite{zheng2015characterization}. As seen from Fig.\ref{fig2}(b), moving the jet along the z axis leads to a noticeable change of the measured intensity of the THz beam, which is a distinguishing feature of Z-scan curves obtained by the known method \cite{sheik1990sensitive}. It is caused by different divergences of the radiation at various positions of the water jet in the caustic, where a nonlinear Kerr lens is induced by the THz radiation field. In view of this we use standard formulas \cite{sheik1990sensitive, yin2000determination, cheung1994optical} to evaluate n$_2$ of water according to the results of our measurements shown in Fig.\ref{fig2}(b):

\begin{equation}
n_2 = {\frac{\Delta T}{0.406I_{in}}\times \frac{\sqrt{2}\lambda}{2\pi L_\alpha (1-2)^{0.25}}}
\label{eq1}
\end{equation}
where $\Delta T$ = 0.013 (Fig.\ref{fig2}(b)) is the difference between the maximum and minimum transmission, $S$ is the linear transmission of the aperture, $L_\alpha$= $\alpha^{-1}$ [1-$\exp(-\alpha L)$] is the effective interaction length, $L$ is the sample thickness, $\alpha$ is the absorption coefficient ($\alpha$ = 100 cm$^{-1}$), $\lambda$ is the wavelength, and $I_{in}$ is the input radiation intensity. The linear transmission of the aperture is 2\%, which allows to maximize the sensitivity of the measurement method but reduces the signal-to-noise ratio. The radiation wavelength was chosen to be $\lambda_0$ = 0.4 mm ($\nu_0$ = 0.75 THz). It corresponds to the maximum of the generation spectrum of the THz radiation (see Fig.\ref{fig1}(b)). The result of the evaluation calculated by formula (\ref{eq1}) gives the value $n_2$ = 3.5$\times$10$^{-10}$ cm$^{2}$/W.

\begin{figure}[htbp]
\centering
\includegraphics[width=7 cm]{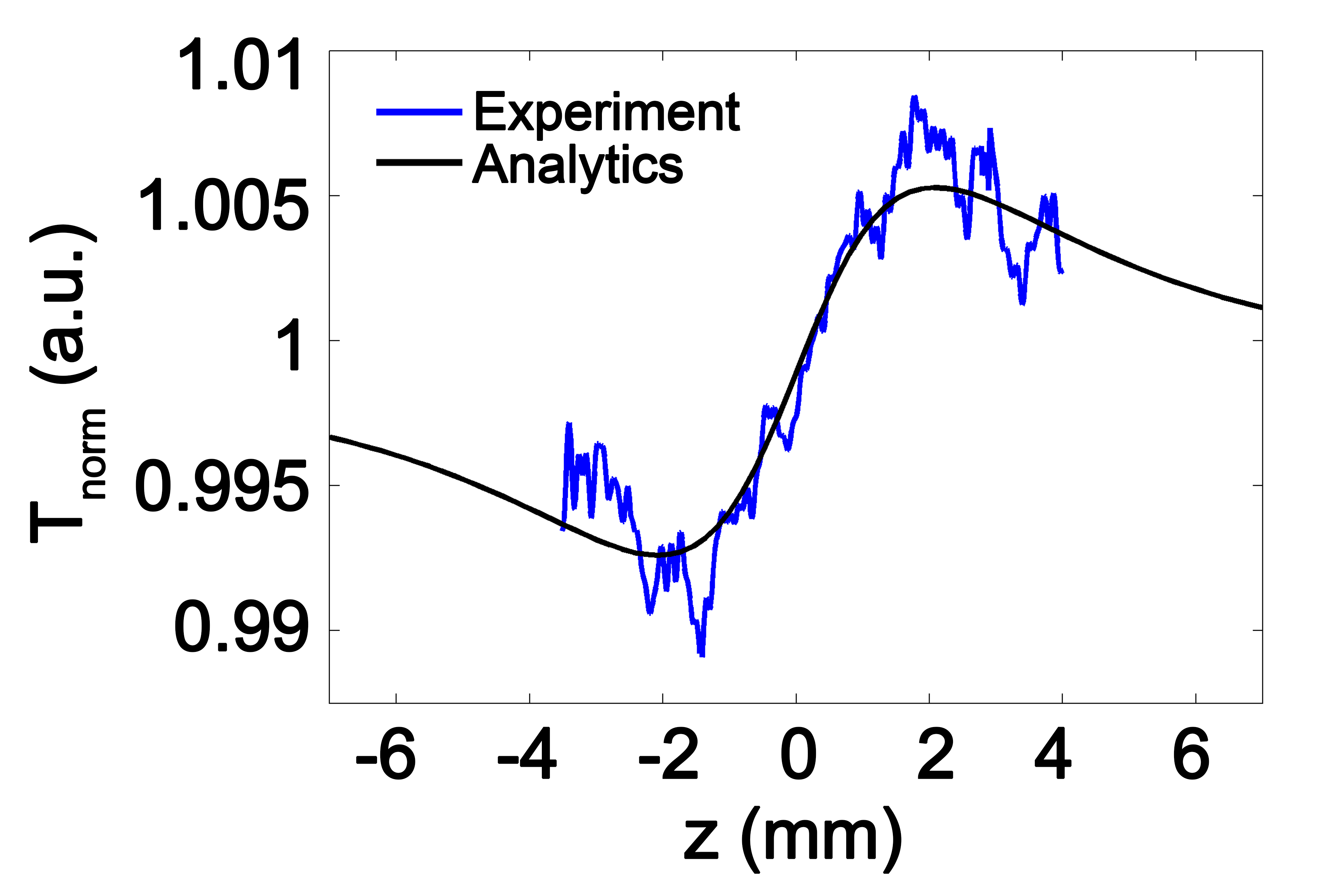}
\caption{Comparison of the experimental results of the closed aperture measurement of Z-scan method for pulsed broadband THz radiation for the water jet 0.1 mm thick with an analytical Z-scan curve for monochromatic radiation with the wavelength of 0.4 mm. The analytical curve was calculated using equation (\ref{eq2}).}
\label{fig3}
\end{figure}

To illustrate the correctness of using formula (1) for the calculation of the nonlinear refractive index coefficient in the case of broadband THz radiation, we compare the experimental data with the analytical Z-scan curve for monochromatic radiation (Fig.\ref{fig3}) calculated by the formula \cite{sheik1990sensitive}:

\begin{equation}
T(z) = {\frac{\int_{-\infty}^{+\infty}P_{T}(\Delta \Phi_{0}(t))dt}{S\int_{-\infty}^{+\infty}P_{i}(t)dt}}
\label{eq2}
\end{equation}
where $P_i$(t)=${\pi w_{0}^2 I_0(t)/2}$ is the instantaneous input power (within the sample), $S$=1-$\exp(-2{r_{a}^2/w_{a}^2})$ is the aperture linear transmittance; the transmitted power through the aperture gives

\begin{equation}
P_{T}(\Delta \Phi_{0}(t))=c\epsilon_0 N_0 \pi \int_{0}^{r_a} a|E_a(r,t)|^2rdr
\label{eq3}
\end{equation}
where 

\begin{equation}
\begin{split}
E_a(r,t)=E(z,r=0,t)\exp(-\alpha L/2)
\times \sum_{m=0}^{+\infty} {\frac{[i \Delta \phi_0 (z,t)]^m}{m!}}{\frac{w_{m0}}{w_m}}\exp({\frac{-r^2}{w_m^2}}{\frac{ikr^2}{2R_m}}+iQ_m)
\label{eq4}
\end{split}
\end{equation}
and $E(z,r=0,t)=E_0sin(2\pi \nu_0 t)w_0/w(z), \Delta \phi_0(z,t)=\Delta \Phi_{0}(t)/(1+z^{2}/z_{0}^{2}), \Delta \Phi_0(t)=k \Delta n_0(t)L.$

The following values are used in these formulas: absorption coefficient $\alpha$ = 100 cm$^{-1}$, sample length $L$ = 0.1 mm, central frequency of the radiation $\nu_0$ = 0.75 THz ($\lambda_0$= 0.4 mm), beam waist radius $w_0$ = 0.5 mm, aperture radius $r_a$ = 1.5 mm, radius of the THz beam $w_a$ = 12.5 mm, intensity of the THz beam in caustic $I_0$ = 10$^8$ W/cm$^2$, nonlinear refractive index coefficient $n_2$ = 3.5$\times$10$^{-10}$ cm$^2$/W. This value of $n_2$ was obtained in the experiment previously.

As can be seen, the experimental Z-scan curve for broadband THz radiation agrees well with the analytical Z-scan curve for monochromatic radiation.

\section{Theoretical estimate of the nonlinear refractive index coefficient}
We estimate the nonlinear refractive index coefficient $n_2$ of liquid water through use of a recent theoretical treatment \cite{dolgaleva2015prediction}. This treatment ascribes the THz nonlinearities in media to a vibrational response that is orders of magnitude larger than typical electronic responses. We make use of Eq. (55) of reference \cite{dolgaleva2015prediction}, which applies to the situation where the THz frequency $\omega_0$ is much smaller than the fundamental vibrational frequency resulting in absorption peak $\lambda$ = 3 $\mu$m ($\omega \approx$ 100 THz) \cite{hale1973optical}. For our experiment this condition is well satisfied, as $\omega_0/2\pi$ is approximately 0.75 THz and $\omega/2\pi$ is 15.9 THz. This formula takes the form:
\begin{equation}
\begin{split}
\bar n_{2,\nu}=\bar n_{2,\nu}^{(1)}+\bar n_{2,\nu}^{(2)}=\frac{3a_1^2m^2\omega^4\alpha_T^2}{32n_0\pi^2q^2N^2k_B^2}\left[ n_{0,\nu}^2-1 \right]^3-\frac{9}{32\pi Nn_0\hslash \omega}\left[ n_{0,\nu}^2-1 \right]^2
\label{eq5}
\end{split}
\end{equation}
We evaluate this expression through use of the following values: $a_1$ is the lattice constant; for our estimations in case of liquid we use the water molecule diameter 2.8$\times$10$^{-8}$ cm \cite{schatzberg1967molecular}, $m$ = 1.6$\times$10$^{-24}$ g is the reduced mass of the vibrational mode, $\alpha_T$ = 0.2$\times$10$^{-3}$ (\textcelsius)$^{-1}$ is the thermal expansion coefficient \cite{kell1967precise}. The parameter $q$ is the effective charge of the chemical bond; for simplicity, we take this quantity to be the electron charge. $N$ is the number density of vibrational units. We calculate this value as ratio between specific gravity of water equal 1 and the total mass of H$_2$O molecule equal to the weight of molecule (1$\times$2 + 16) times amu (1.67$\times$10$^{-24}$). It results in $N$  = 3.3$\times$10$^{22}$ in 1 cm$^3$. We  take the refractive index as $n_{0,\nu}$ = 2.3 which is the averaged refractive index in 0.3--1.0 THz region \cite{thrane1995thz}. Using these values, we find that the predicted value of $n_2$ for water in the low-frequency limit is $n_2$ = 5.0$\times$10$^{-10}$ cm$^2$/W.

\section{Summary}
In conclusion we have experimentally demonstrated the possibility of direct measurement of the nonlinear refractive index coefficient $n_2$ of water in the THz frequency range. Z-scan curves experimentally obtained for broadband THz radiation are in good agreement with the analytical model of the method for monochromatic radiation. The value of the nonlinear refractive index coefficient of water calculated from the experiment is $n_{2}$ = 3.5$\times$10$^{-10}$ cm$^{2}$/W, which is 6 orders of magnitude higher than for the visible and IR ranges \cite{paillette1969recherches, smith1977superbroadening, ho1979optical,boyd2003nonlinear} where $n_{2}$ has the magnitude of $10^{-16}$ cm$^{2}$/W. These results demonstrate the high cubic nonlinearity of water in the THz frequency range and confirm a recent theoretical prediction \cite{dolgaleva2015prediction} that the ionic vibrational contribution to the third-order susceptibility makes renders THz nonlinearities much larger than typical optical-frequency nonlinearities. Therefore, in terms of applications, our demonstration opens up new perspectives for studying various materials in the THz range. Nonlinear optics, in its turn, finds applications in the creation of light modulators, transistors, switchers, etc. in this spectral region.

\section*{Funding}

Government of the Russian Federation (08-08); Ministry of Education and Science of the Russian Federation (Minobrnauka) (3.9041.2017/7.8); USA Army Research Office (W911NF-17-1-0428); Russian Foundation of Basic Research (RFBR) (19-02-00154 A).




\begin{thebibliography}{10}
\newcommand{\enquote}[1]{``#1''}

\bibitem{baierl2016nonlinear}
S.~Baierl, M.~Hohenleutner, T.~Kampfrath, A.~Zvezdin, A.~Kimel, R.~Huber, and
  R.~Mikhaylovskiy, \enquote{Nonlinear spin control by terahertz-driven
  anisotropy fields,} {\protect\JournalTitle{Nature Photonics}} \textbf{10},
  715 (2016).

\bibitem{dong2017global}
J.~Dong, A.~Locquet, M.~Melis, and D.~Citrin, \enquote{Global mapping of
  stratigraphy of an old-master painting using sparsity-based terahertz
  reflectometry,} {\protect\JournalTitle{Scientific reports}} \textbf{7}
  (2017).

\bibitem{kaltenecker2016ultrabroadband}
K.~J. Kaltenecker, A.~Tuniz, S.~C. Fleming, A.~Argyros, B.~T. Kuhlmey,
  M.~Walther, and B.~M. Fischer, \enquote{Ultrabroadband perfect imaging in
  terahertz wire media using single-cycle pulses,}
  {\protect\JournalTitle{Optica}} \textbf{3}, 458--464 (2016).

\bibitem{nagatsuma2016advances}
T.~Nagatsuma, G.~Ducournau, and C.~C. Renaud, \enquote{Advances in terahertz
  communications accelerated by photonics,} {\protect\JournalTitle{Nature
  Photonics}} \textbf{10}, 371 (2016).

\bibitem{grachev2018wireless}
Y.~V. Grachev, X.~Liu, S.~E. Putilin, A.~N. Tsypkin, V.~G. Bespalov, S.~A.
  Kozlov, and X.-C. Zhang, \enquote{Wireless data transmission method using
  pulsed thz sliced spectral supercontinuum,} {\protect\JournalTitle{IEEE
  Photonics Technology Letters}} \textbf{30}, 103--106 (2018).

\bibitem{bowlan2017probing}
P.~Bowlan, J.~Bowlan, S.~Trugman, R.~V. Aguilar, J.~Qi, X.~Liu, J.~Furdyna,
  M.~Dobrowolska, A.~Taylor, D.~Yarotski \emph{et~al.}, \enquote{Probing and
  controlling terahertz-driven structural dynamics with surface sensitivity,}
  {\protect\JournalTitle{Optica}} \textbf{4}, 383--387 (2017).

\bibitem{smolyanskaya2018glycerol}
O.~Smolyanskaya, I.~Schelkanova, M.~Kulya, E.~Odlyanitskiy, I.~Goryachev,
  A.~Tcypkin, Y.~V. Grachev, Y.~G. Toropova, and V.~Tuchin, \enquote{Glycerol
  dehydration of native and diabetic animal tissues studied by thz-tds and nmr
  methods,} {\protect\JournalTitle{Biomedical optics express}} \textbf{9},
  1198--1215 (2018).

\bibitem{zhang2017extreme}
X.~C. Zhang, A.~Shkurinov, and Y.~Zhang, \enquote{Extreme terahertz science,}
  {\protect\JournalTitle{Nature Photonics}} \textbf{11}, 16 (2017).

\bibitem{shalaby2015demonstration}
M.~Shalaby and C.~P. Hauri, \enquote{Demonstration of a low-frequency
  three-dimensional terahertz bullet with extreme brightness,}
  {\protect\JournalTitle{Nature communications}} \textbf{6}, 5976 (2015).

\bibitem{gaal2006nonlinear}
P.~Gaal, K.~Reimann, M.~Woerner, T.~Elsaesser, R.~Hey, and K.~H. Ploog,
  \enquote{Nonlinear terahertz response of n-type gaas,}
  {\protect\JournalTitle{Physical review letters}} \textbf{96}, 187402 (2006).

\bibitem{hebling2010observation}
J.~Hebling, M.~C. Hoffmann, H.~Y. Hwang, K.-L. Yeh, and K.~A. Nelson,
  \enquote{Observation of nonequilibrium carrier distribution in ge, si, and
  gaas by terahertz pump--terahertz probe measurements,}
  {\protect\JournalTitle{Physical Review B}} \textbf{81}, 035201 (2010).

\bibitem{hoffmann2009thz}
M.~C. Hoffmann, J.~Hebling, H.~Y. Hwang, K.-L. Yeh, and K.~A. Nelson,
  \enquote{Thz-pump/thz-probe spectroscopy of semiconductors at high field
  strengths,} {\protect\JournalTitle{JOSA B}} \textbf{26}, A29--A34 (2009).

\bibitem{sharma2012carrier}
G.~Sharma, I.~Al-Naib, H.~Hafez, R.~Morandotti, D.~Cooke, and T.~Ozaki,
  \enquote{Carrier density dependence of the nonlinear absorption of intense
  thz radiation in gaas,} {\protect\JournalTitle{Optics express}} \textbf{20},
  18016--18024 (2012).

\bibitem{turchinovich2012self}
D.~Turchinovich, J.~M. Hvam, and M.~C. Hoffmann, \enquote{Self-phase modulation
  of a single-cycle terahertz pulse by nonlinear free-carrier response in a
  semiconductor,} {\protect\JournalTitle{Physical Review B}} \textbf{85},
  201304 (2012).

\bibitem{li2015terahertz}
S.~Li, G.~Kumar, and T.~E. Murphy, \enquote{Terahertz nonlinear conduction and
  absorption saturation in silicon waveguides,} {\protect\JournalTitle{Optica}}
  \textbf{2}, 553--557 (2015).

\bibitem{baierl2016terahertz}
S.~Baierl, J.~H. Mentink, M.~Hohenleutner, L.~Braun, T.-M. Do, C.~Lange,
  A.~Sell, M.~Fiebig, G.~Woltersdorf, T.~Kampfrath \emph{et~al.},
  \enquote{Terahertz-driven nonlinear spin response of antiferromagnetic nickel
  oxide,} {\protect\JournalTitle{Physical review letters}} \textbf{117}, 197201
  (2016).

\bibitem{vicario2017subcycle}
C.~Vicario, M.~Shalaby, and C.~P. Hauri, \enquote{Subcycle extreme
  nonlinearities in gap induced by an ultrastrong terahertz field,}
  {\protect\JournalTitle{Physical review letters}} \textbf{118}, 083901 (2017).

\bibitem{lin2018measurement}
S.~Lin, S.~Yu, and D.~Talbayev, \enquote{Measurement of quadratic terahertz
  optical nonlinearities using second-harmonic lock-in detection,}
  {\protect\JournalTitle{Physical Review Applied}} \textbf{10}, 044007 (2018).

\bibitem{kaur2010terahertz}
G.~Kaur, P.~Han, and X.~Zhang, \enquote{Terahertz induced nonlinear effects in
  doped silicon observed by open-aperture z-scan,} in \emph{Infrared Millimeter
  and Terahertz Waves (IRMMW-THz), 2010 35th International Conference on,}
  (IEEE, 2010), pp. 1--2.

\bibitem{dolgaleva2015prediction}
K.~Dolgaleva, D.~V. Materikina, R.~W. Boyd, and S.~A. Kozlov,
  \enquote{Prediction of an extremely large nonlinear refractive index for
  crystals at terahertz frequencies,} {\protect\JournalTitle{Physical Review
  A}} \textbf{92}, 023809 (2015).

\bibitem{woldegeorgis2018thz}
A.~Woldegeorgis, T.~Kurihara, B.~Beleites, J.~Bossert, R.~Grosse, G.~G. Paulus,
  F.~Ronneberger, and A.~Gopal, \enquote{Thz induced nonlinear effects in
  materials at intensities above 26 gw/cm 2,} {\protect\JournalTitle{Journal of
  Infrared, Millimeter, and Terahertz Waves}} \textbf{39}, 667--680 (2018).

\bibitem{Tcypkin2018}
A.~Tcypkin, S.~Putilin, M.~Kulya, M.~Melnik, A.~Drozdov, V.~Bespalov, X.-C.
  Zhang, R.~Boyd, and S.~Kozlov, \enquote{Experimental estimate of the
  nonlinear refractive index of crystalline znse in the terahertz spectral
  range,} {\protect\JournalTitle{Bulletin of the Russian Academy of Sciences:
  Physics}} \textbf{82}, 1547--1549 (2018).

\bibitem{jin2017observation}
Q.~Jin, Y.~E, K.~Williams, J.~Dai, and X.-C. Zhang, \enquote{Observation of
  broadband terahertz wave generation from liquid water,}
  {\protect\JournalTitle{Applied Physics Letters}} \textbf{111}, 071103 (2017).

\bibitem{doi:10.1063/1.5054599}
Y.~E, Q.~Jin, A.~Tcypkin, and X.-C. Zhang, \enquote{Terahertz wave generation
  from liquid water films via laser-induced breakdown,}
  {\protect\JournalTitle{Applied Physics Letters}} \textbf{113}, 181103 (2018).

\bibitem{paillette1969recherches}
M.~Paillette, \enquote{Recherches exp{\'e}rimentales sur les effets kerr
  induits par une onde lumineuse,} in \emph{Annales de Physique,}  vol.~14
  (1969), pp. 671--712.

\bibitem{smith1977superbroadening}
W.~L. Smith, P.~Liu, and N.~Bloembergen, \enquote{Superbroadening in h 2 o and
  d 2 o by self-focused picosecond pulses from a yalg: Nd laser,}
  {\protect\JournalTitle{physical Review A}} \textbf{15}, 2396 (1977).

\bibitem{ho1979optical}
P.~Ho and R.~Alfano, \enquote{Optical kerr effect in liquids,}
  {\protect\JournalTitle{Physical Review A}} \textbf{20}, 2170 (1979).

\bibitem{boyd2003nonlinear}
R.~W. Boyd, \emph{Nonlinear optics} (Elsevier, 2003).

\bibitem{yang1971generation}
K.~Yang, P.~Richards, and Y.~Shen, \enquote{Generation of far-infrared
  radiation by picosecond light pulses in linbo3,}
  {\protect\JournalTitle{Applied Physics Letters}} \textbf{19}, 320--323
  (1971).

\bibitem{watanabe1989new}
A.~Watanabe, H.~Saito, Y.~Ishida, M.~Nakamoto, and T.~Yajima, \enquote{A new
  nozzle producing ultrathin liquid sheets for femtosecond pulse dye lasers,}
  {\protect\JournalTitle{Optics Communications}} \textbf{71}, 301--304 (1989).

\bibitem{kislin2018self}
D.~A. Kislin, M.~A. Knyazev, Y.~A. Shpolyanskii, and S.~A. Kozlov,
  \enquote{Self-action of nonparaxial few-cycle optical waves in dielectric
  media,} {\protect\JournalTitle{JETP Letters}} \textbf{107}, 753--760 (2018).

\bibitem{zheng2015characterization}
X.~Zheng, R.~Chen, G.~Shi, J.~Zhang, Z.~Xu, T.~Jiang \emph{et~al.},
  \enquote{Characterization of nonlinear properties of black phosphorus
  nanoplatelets with femtosecond pulsed z-scan measurements,}
  {\protect\JournalTitle{Optics letters}} \textbf{40}, 3480--3483 (2015).

\bibitem{sheik1990sensitive}
M.~Sheik-Bahae, A.~A. Said, T.-H. Wei, D.~J. Hagan, and E.~W. Van~Stryland,
  \enquote{Sensitive measurement of optical nonlinearities using a single
  beam,} {\protect\JournalTitle{IEEE journal of quantum electronics}}
  \textbf{26}, 760--769 (1990).

\bibitem{yin2000determination}
M.~Yin, H.~Li, S.~Tang, and W.~Ji, \enquote{Determination of nonlinear
  absorption and refraction by single z-scan method,}
  {\protect\JournalTitle{Applied Physics B}} \textbf{70}, 587--591 (2000).

\bibitem{cheung1994optical}
Y.~Cheung and S.~Gayen, \enquote{Optical nonlinearities of tea studied by
  z-scan and four-wave mixing techniques,} {\protect\JournalTitle{JOSA B}}
  \textbf{11}, 636--643 (1994).

\bibitem{hale1973optical}
G.~M. Hale and M.~R. Querry, \enquote{Optical constants of water in the 200-nm
  to 200-$\mu$m wavelength region,} {\protect\JournalTitle{Applied optics}}
  \textbf{12}, 555--563 (1973).

\bibitem{schatzberg1967molecular}
P.~Schatzberg, \enquote{Molecular diameter of water from solubility and
  diffusion measurements,} {\protect\JournalTitle{The Journal of Physical
  Chemistry}} \textbf{71}, 4569--4570 (1967).

\bibitem{kell1967precise}
G.~Kell, \enquote{Precise representation of volume properties of water at one
  atmosphere,} {\protect\JournalTitle{Journal of Chemical and Engineering
  data}} \textbf{12}, 66--69 (1967).

\bibitem{thrane1995thz}
L.~Thrane, R.~H. Jacobsen, P.~U. Jepsen, and S.~Keiding, \enquote{Thz
  reflection spectroscopy of liquid water,} {\protect\JournalTitle{Chemical
  Physics Letters}} \textbf{240}, 330--333 (1995).

\end{thebibliography}






\end{document}